# Determination of the electromechanical limits of high-performance Nb3Sn Rutherford cables under transverse stress from a single-wire experiment

L. Gamperle [a], J. Ferradas [a,b], C. Barth [a], B. Bordini [b], D. Tommasini [b], and C. Senatore [a]

[a] Department of Quantum Matter Physics (DQMP), University of Geneva, 1211 Geneva, Switzerland
[b] CERN, Geneva, Switzerland

**Abstract**

The development of high-field accelerator magnets capable of providing 16 T dipolar fields is an indispensable technological breakthrough needed for the 100 TeV energy-frontier targeted by the Future Circular Collider (FCC). As these magnets will be based on Nb3Sn Rutherford cables, the degradation of the conductor performance due to the large electro-magnetic stresses becomes a parameter with a profound impact on the magnet design. In this work, we investigated the stress dependence and the irreversible reduction of the critical current under compressive transverse load in high performance Powder-In-Tube (PIT) Nb3Sn wires. Tests were performed in magnetic fields ranging between 16 T and 19 T on wires that were resin-impregnated similarly to the wires in the Rutherford cables of accelerator magnets. The scope was to predict the degradation of the cable under stress from a single-wire experiment. Interestingly, the irreversible stress limit, $\sigma_{irr}$, defined as the stress level corresponding to a permanent reduction of the critical current by 5% with respect to its initial value, was found to depend on the applied magnetic field. This observation allowed us to shed light on the mechanism dominating the irreversible reduction of the wire performance and to compare and reconcile our results with the irreversible limits measured on Rutherford cables, typically tested at fields below 12 T.

**Introduction**

The goal of increasing the potential of discovering new physics is pushing the High Energy Physics community to conceive novel experiments based on a highest-energy hadron collider with a centre-of-mass collision energy of 100 TeV [1]. One of the main challenges in view of this Future Circular Collider (FCC) is the development of high-field superconducting accelerator magnets. Assuming a ring circumference of 100 km, the dipole field needed to reach 100 TeV must be about 16 T, as the energy at the collision is given by

$$E[\text{TeV}] = 3 \cdot 10^{-4} \times B \times R, \tag{1}$$

where B is the intensity of the magnetic field and R is the radius of the ring [2], both in SI units. To reach a field level that is almost twice that of the magnets installed in the Large Hadron Collider (LHC), the dipole magnets for the FCC rely on niobium-tin (Nb3Sn) superconductors. Over the last two decades, high-field Nb3Sn magnet technology has made great progresses thanks to the ITER conductor development [3], R&D programs co-funded by the European Commission [4], and US-LARP [5]. Moreover, the High-Luminosity upgrade of the LHC (HL-LHC) will include two pairs of dipoles operating at 11 T and 24 quadrupoles based on Nb3Sn. These will be the first Nb3Sn magnets ever operating in a particle accelerator and thus represent an important test bench in the perspective of the FCC developments.

A 16 T field in a dipole configuration translates into a requirement of a minimum critical current density, $J_c$, of more than 1500 A/mm$^2$ in the superconductor at 16 T and 4.2 K [6], which is a target currently beyond state-of-the-art for industrial Nb3Sn wires [7, 8]. Reaching this target requires work on novel methods such as grain refinement and artificial pinning centers [9,10], with some promising results already achieved [11]. In order to be suitable for the 16 T dipoles, Nb3Sn wires must meet some other essential requirements [12]: (i) a good thermal and electrical stabilization, provided by an adequate amount of high purity Cu, (ii) a low superconducting filament size, to achieve the necessary field quality and (iii) good tolerance to the mechanical loads imposed when assembling the wires in Rutherford cables and by the electromagnetic forces during operation.

The degradation of the conductor performance due to stress is a parameter with a profound impact on the magnet design. In the frame of the European project EuroCirCol, four options are being studied for the dipoles of the FCC: cos-theta [13], block-type [14], common-coil [15] and canted cos-theta [16]. All present designs entail peak stresses in the 150-200 MPa range on the Nb$_3$Sn Rutherford cables. Typically, the stress on the conductor at the nominal field is compressive in the transverse direction and localized at the magnet midplane. Transverse stress induces a large reduction of the critical current, I$_c$, of multifilamentary Nb$_3$Sn wires [17]. The I$_c$ reduction has a reversible component, which is fully recovered when removing the load and is associated to the reversible reduction under stress of the upper critical field [18], and an irreversible and thus permanent component. Two phenomena work together to determine the irreversible reduction of the wire performance: cracks and plastic deformations [19]. First, Nb$_3$Sn is a brittle intermetallic compound characterized by a strong propensity to fracture. Secondly, the superconducting filaments in a Nb$_3$Sn wire are embedded in a soft Cu matrix; hence it is very keen to deform plastically. Plastic deformation of the stabilizing matrix due to an external load leaves the Nb$_3$Sn filaments under residual stress after unload.

Assessing the stress tolerance of Nb$_3$Sn Rutherford cables becomes therefore essential to estimate the transport current properties of the conductor within the magnet. However, experiments with full-size cables under transverse compressive stress are difficult to conduct and are possible only in very few facilities around the world. At the University of Twente cables can be tested under high transverse compression ($\geq$ 200 MPa) [20] in an external magnetic field up to 11 T, the major limitation being that the load can be applied on a reduced sample length (approximately 5 cm). CERN has developed a sample holder for the 10 T FRESCA dipole magnet [21] to test cables under transverse force, with the stress applied uniformly over a 70 cm long region of the sample. The results of a measurement campaign up to 160 MPa were reported recently in Ref. [22]. A complication of this setup is that samples need to be warmed up in order to change the pressure, as the load is exerted using a bladder-and-key structure [21]. Similar electromechanical experiments can be performed in two more test stations. At NHMFL it is possible to test Rutherford cables at stresses up to 200 MPa in a 13 T split-coil magnet [23]. At Fermilab there is a device that allows measuring I$_c$ under transverse compression of a single superconducting wire housed in a dummy cable up to 200 MPa, 14 T [24]. In both cases, stress is applied over a length of few centimeters, 12 cm at NHMFL and 6 cm at Fermilab.

This paper presents the results of an activity of EuroCirCol on the electromechanical properties of Nb$_3$Sn wires under transverse loads. The scope was to extract from a single-wire experiment quantitative information about the degradation of a Rutherford cable under transverse stress. We investigated the stress dependence and the irreversible reduction of the critical current under transverse compressive load in high-performance Powder-In-Tube (PIT) Nb$_3$Sn wires. Tests were performed on resin-impregnated single wires at 4.2 K and at various magnetic fields between 16 T and 19 T, using a modified Walters spring probe [25, 26]. The measurement campaign focused on two aspects: the influence of the type of impregnation on the irreversible stress limit of the wire and the identification of the main mechanism responsible for the irreversible reduction of I$_c$.

**Experimental details**

The PIT Nb$_3$Sn wire tested during this measurement campaign has 192 filaments embedded in a high-purity Cu matrix, and the resulting Cu/non-Cu volume ratio is approximately 1.22. Tests were performed at two different diameters, 1.00 mm and 0.85 mm. Both were developed by Bruker EAS for CERN: the one at 1.00 mm for the FRESCA2 dipole [27], and the one at 0.85 mm for the new main ring inner triplet quadrupoles, identified by the acronym MQXF, in the high luminosity LHC upgrade [28]. Samples from different billets and with different reaction schedules were examined to consolidate the results of the study; their main characteristics are reported in table I.

**Table I** *Main characteristics of the investigated PIT Nb$_3$Sn wires*

| Billet ID | #31712 | #14310 | #29992 |
|---|---|---|---|
| Wire diameter | 1.00 | 1.00 | 0.85 |
| Heat treatment schedule | 620°C/100h +640°C/90h | 620°C/120h +650°C/90h | 600°C/100h +625°C/200h |
| I$_c$(16 T) [A] | 337 | 379 | 239 |
| I$_c$(19 T) [A] | 164 | 146 | 102 |

The probe used for this measurement campaign is described in details in [19,25-26]. It is based on a one-turn Walters spring, which has been cut parallel to the wire sample into two parts: the lower part can move axially against the fixed upper part, transmitting a transverse force to the sample. The wire is confined in a U-shaped groove, whose width is adapted to the sample diameter, and the upper part of the spiral hosts an anvil designed to fit the groove. To be representative of the wires in Rutherford cables, the sample is resin-impregnated. The gauge length is 126 mm, which is significantly longer than the typical twist pitch length of Nb$_3$Sn wires, and this allows applying stringent voltage criteria like 0.1 µV/cm or lower. Current transfer to the sample is assured by soldering the superconducting wire onto copper terminals over about 250 mm.

**Results and discussion**

Fig. 1(a) shows the results of a measurement performed at T = 4.2 K and B = 19 T on a wire sample from billet #31712 impregnated with a mixture of epoxy resin (type L + hardener L provided by R&G Faserverbundenwerkstoffe) and filler (thixotropic agent provided by SCS-Füllstoffe) in a 100:40:2 weight ratio. Flexural tests performed at CERN [29,30] revealed that the elastic modulus of loaded epoxy type-L is comparable to the one of CDT-101K, which is the resin used for the impregnation of large coils for accelerator magnets. In our tests, I$_c$ is measured for increasing values of the applied transverse force, up to 35 kN, and the stress on the wire is calculated as the force divided by the groove area. The solid squares in the graph show the dependence on applied stress, $\sigma$, of the reduced I$_c^{load}$, i.e. I$_c$ under load normalized to the critical current at zero applied stress, I$_{c0}$. Each open square in Fig. 1(a) represents the fraction of I$_{c0}$ recovered after unloading the stress from the value indicated on the x-axis to zero, I$_c^{unload}(\sigma \to 0)$. A reduction of 5% with respect to I$_{c0}$ is considered here as the onset of the irreversible behavior and the corresponding stress is indicated as the irreversible stress limit, $\sigma_{irr}$. The wire impregnated with epoxy type L reaches its irreversible limit at $\sigma_{irr}$ = 110 MPa, which is well below the peak stress of 150-200 MPa of the FCC 16 T dipole designs. $\sigma_{irr}$ is substantially higher when the wire is embedded in a reinforced impregnation. We repeated the same test as in Fig. 1(a) using two different impregnation schemes. Fig. 1(b) reports the results of a measurement performed on a wire sample from billet #31712 impregnated with Stycast, whose elastic modulus is about the double compared to CDT-101K and epoxy type-L [29,30]. The wire tested in Fig. 1(c), which belongs to the billet #14310, was inserted in a glass-fiber sleeve and impregnated with the regular mixture of epoxy type L and filler. As the diameter of the wire with its sleeve is about 1.1 mm, the measurement was performed using a probe head with a slightly larger groove compared to that used in the experiments shown in Fig. 1(a) and (b) (1.30 mm vs. 1.15 mm). The straightforward conclusion is that adding rigidity to the impregnation moves the onset of the irreversible reduction of the wire critical current towards higher stresses. In these measurements, performed at 4.2 K, 19 T, $\sigma_{irr}$ is increased from 110 MPa, when using epoxy type L impregnation, to about 160 MPa both with Stycast and glass-fiber reinforcement.

Interestingly, when plotting the normalized critical current at a given applied stress, I$_c^{load}(\sigma)$/I$_{c0}$, versus the normalized critical current measured after the corresponding stress unload, I$_c^{unload}(\sigma \to 0)$/I$_{c0}$, measurements exhibit a scaling behavior. This is shown in Fig. 2, where the data from Fig. 1 are replotted along with the results of a test on the 0.85 mm-wire from billet #29992 and the data on a PIT wire with the same filament layout extracted from Ref. [19]. The value of the applied load that determines a given reduction of I$_c^{load}$ depends on the type of impregnation. However, when a given reduction of I$_c^{load}$ is measured, this determines also the level of irreversible reduction of I$_c^{unload}$, regardless of the type of impregnation. The origin of the scaling behavior is thus related to the intrinsic response of Nb$_3$Sn filaments to the local stress: the differences observed in the measured values of $\sigma_{irr}$ are solely linked to the redistribution of stress between the superconductor and the medium that transmits the load.

As reported already in the introduction, the irreversible reduction of the critical current under transverse stress results from the combination of the effects of plastic deformation of the Cu matrix and cracks in the Nb$_3$Sn filaments. However, the effects of these two phenomena on the measured critical current are different. The residual stress imposed from the plastically deformed matrix to Nb$_3$Sn determines a permanent distortion of its crystal structure. This in turn induces a reduction of the upper critical field after unload, B$_{c2}^{unload}$, with respect to the value measured on the virgin wire, B$_{c2,0}$. For a wide range of multifilamentary Nb$_3$Sn superconductors the scaling law for the field dependence of the critical current is well approximated by the expression [31]:

$$I_c(B) = C \left(\frac{B}{B_{c2}}\right)^{-0.5} \left(1 - \frac{B}{B_{c2}}\right)^2 \qquad (2)$$

where C is a prefactor almost independent of σ over a large range of applied transverse stresses [21]. It follows from eq. (2) that $I_c^{unload}/I_{c0}$ has to be a function of the applied magnetic field when the irreversible reduction of $I_c^{unload}$ is driven by the reduction of the upper critical field due to the plastic deformation of the matrix. In this scenario, also $\sigma_{irr}$, which is defined as the value of the applied stress corresponding to $I_c^{unload}/I_{c0} = 0.95$, is expected to depend on B. On the other hand, filament cracks generate a permanent reduction of the current carrying cross section and, thus, the resulting $I_c^{unload}/I_{c0}$ and $\sigma_{irr}$ are independent of the magnetic field.

In order to assess the dominant mechanism behind the irreversible reduction of $I_c^{unload}$ observed in our experiments, we tested at B = 16 T the dependence on transverse stress of $I_c$ on a new sample from billet #31712 and impregnated with epoxy type L, labelled as #31712_bis. In addition, we measured the field dependence of $I_c^{load}$ and $I_c^{unload}$ at given values of the applied stress, namely σ = 76, 104, 135, 173 and 207 MPa. This allowed us to determine the stress dependence of the upper critical field under load, $B_{c2}^{load}(\sigma)$, and after unload, $B_{c2}^{unload}(\sigma \to 0)$, via the well-known Kramer extrapolation [32]; values are reported in table II. In Fig. 3 the data measured on sample #31712_bis at 16 T and 19 T are plotted together with the curves from Fig. 1(a). The direct comparison of the two datasets at 19 T is shown to witness the good reproducibility of the experiment. The lower stress sensitivity under transverse load of $I_c$ at 16 T is a straightforward consequence of eq. (2) with $B_{c2} = B_{c2}^{load}(\sigma)$. However, the most important result of Fig. 3 comes from the comparison of the unload curves. As expected in a scenario of irreversible reduction of $I_c^{unload}$ dominated by the plastic deformation of the Cu matrix, $\sigma_{irr}$ is found to depend on the magnetic field. Its value at 16 T is 135 MPa, i.e. 25 MPa higher compared to the value measured at 19 T.

**Table II** *Transverse stress dependence of the upper critical field under load, $B_{c2}^{load}$, and after unload, $B_{c2}^{unload}$ for the 1-mm PIT wire #31712, as determined from the Kramer extrapolation.*

| Transverse stress [MPa] | $B_{c2}^{load}$ [T] | $B_{c2}^{unload}$ [T] |
|---|---|---|
| 76 | 24.9 | 25.8 |
| 104 | 24.4 | 25.6 |
| 135 | 23.9 | 25.6 |
| 173 | 23.3 | 25.0 |
| 207 | 22.8 | 24.9 |

As the facilities developed to study the effect of transverse stress on the critical current of full-size Rutherford cables are limited to fields below or equal to 13 T [20-23], a direct comparison of the irreversible stress limit extracted from these measurements at lower field with our results obtained on a single wire at B = 16 T or 19 T would be misleading. Nevertheless, we demonstrate in the following that it is possible to reconcile the data obtained on cables and those measured on wires. When the irreversible reduction of the critical current is dominated by plastic deformation phenomena, the field and stress dependences of the ratio between $I_c^{unload}(\sigma \to 0)$ and $I_c^{load}(\sigma)$ are given by the expression:

$$\frac{I_c^{unload}}{I_c^{load}}(B,\sigma) = \left[\frac{B_{c2}^{load}(\sigma)}{B_{c2}^{unload}(\sigma)}\right]^{\frac{3}{2}} \left[\frac{B_{c2}^{unload}(\sigma)-B}{B_{c2}^{load}(\sigma)-B}\right]^2, \qquad (3)$$

that follows directly from eq. (2). Fig. 4 reports the stress dependence of $I_c^{unload}/I_c^{load}$ from three different datasets. The curves at B = 16 T and B = 19 T correspond to the data reported in Fig. 3 and Fig. 1(a), respectively. The third dataset comes from the test of an 18-wire Rutherford cable, made with the same 1 mm-diameter PIT wire of our tests and impregnated with CDT-101K, performed at a peak field of 11.4 T at the University of Twente [33]. Interestingly, the trend of the experimental data for the dependences on applied stress and magnetic field are well reproduced by eq. (3), whose predictions based on the $B_{c2}$ data from table II are represented in the plot as open stars. This analysis proves that the results of the experiments on a single wire are consistent with those performed on a full-size cable. The main reasons why this conclusion was a priori unexpected are two: (i) when a Rutherford cable is assembled and compacted, wires are substantially deformed; (ii) under a transverse load, stress is redistributed non-uniformly among the wires that compose the cable. In spite of that, the simplified load geometry of our experiment on a round wire is able to capture the key aspects of the response to transverse stress of a Rutherford cable.

**Conclusions**

In this work, we have investigated the stress dependence and the irreversible reduction of the critical current under compressive transverse load in high-performance PIT $Nb_3Sn$ wires. Tests were performed on single wires that were resin-impregnated similarly to the wires in the Rutherford cables of accelerator magnets. The main results of the present study are the following:
- The irreversible stress limit of the wire is found to be largely controlled by the rigidity of the impregnation. In our tests, we measured an increase of $\sigma_{irr}$ of about 50 MPa by replacing a soft epoxy-impregnation with Stycast or by adding a glass-fiber reinforcement.
- A scaling behavior is observed when the normalized critical current at a given applied stress, $I_c(\sigma)/I_{c0}$ is plotted against the normalized critical current measured after the corresponding stress unload, $I_c^{unload}(\sigma \rightarrow 0)/I_{c0}$, regardless of the type of impregnation. We identified the origin of this scaling behavior in the intrinsic response of $Nb_3Sn$ filaments to the local stress.
- The comparison of $I_c(\sigma)$ measurements performed at B = 16 T and B = 19 T shows that the fraction of critical current recovered after unload, $I_c^{unload}(\sigma \rightarrow 0)/I_{c0}$, depends on the magnetic field. From this observation we concluded that the irreversible reduction of $I_c^{unload}$ is dominated by the residual stress in the $Nb_3Sn$ filaments, which arises from the plastic deformation of the Cu matrix under load.
- Finally, we have shown that the results of the tests under transverse stress on a single wire are fully consistent with those performed on a full-size Rutherford cable and that is therefore possible to extract quantitative information about the degradation of the cable from a single-wire experiment.

**Acknowledgements**

The authors acknowledge Damien Zurmuehle for his technical support with the electro-mechanical measurements and Tommaso Bagni for the useful discussions. This work was supported in part by the European Union's Horizon 2020 research and innovation programme under Grant 654305, EuroCirCol project.

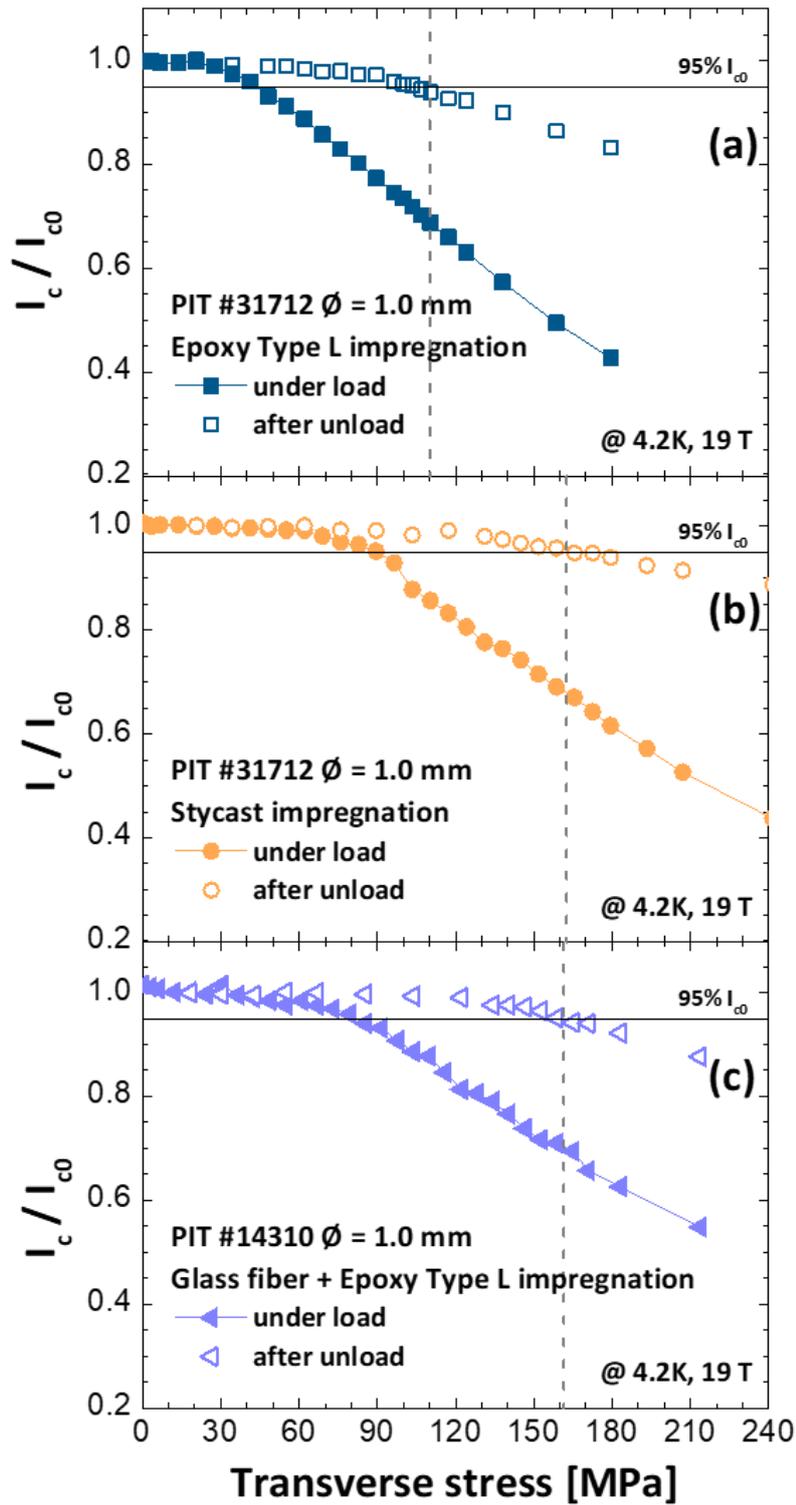

**Figure 1** Dependence on the applied transverse stress at T = 4.2 K, B = 19 T of the critical current $I_c$ normalized to the critical current at zero applied stress, $I_{c0}$, for the 1-mm PIT wire (a) bare and impregnated with epoxy type-L; (b) bare and impregnated with Stycast; (c) in a glass-fiber sleeve and impregnated with epoxy type-L. Solid and open symbols correspond to the measurement under load and after unload, respectively. Dashed lines indicate the irreversible stress limits.

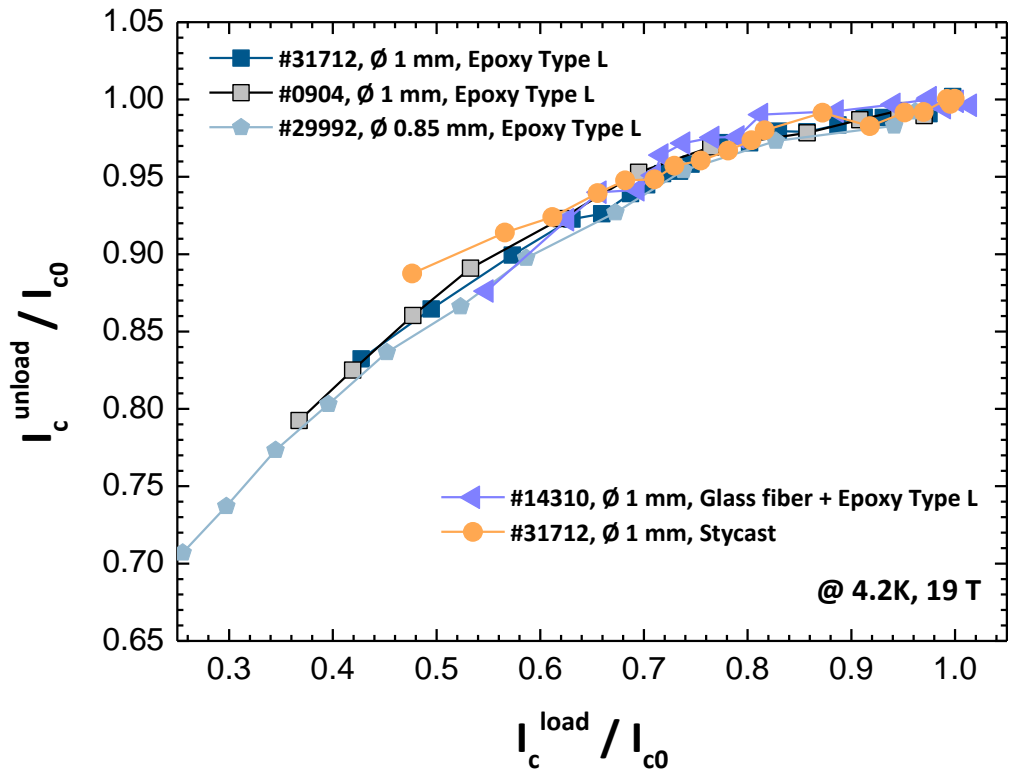

**Figure 2** Scaling behavior of the normalized critical current under load, $I_c^{load}/I_{c0}$, plotted against the normalized critical current after the corresponding stress unload, $I_c^{unload}/I_{c0}$. The reported data were collected at T = 4.2 K, B = 19 T during five distinct tests performed on PIT wires from various billets and using different impregnation schemes.

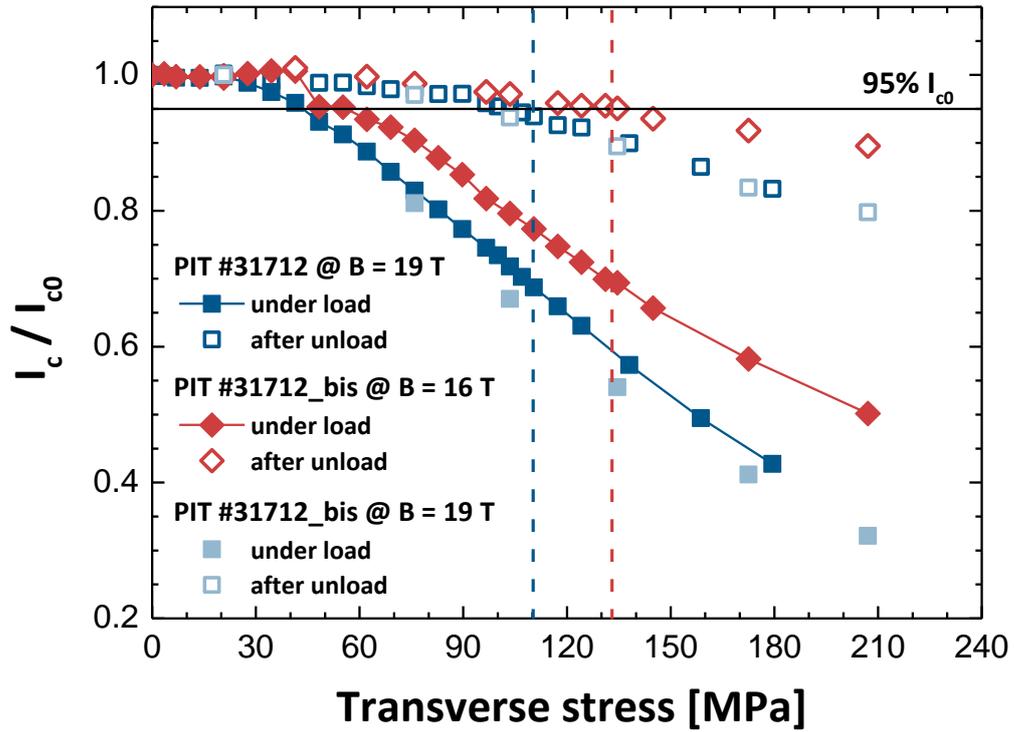

**Figure 3** Dependence on the applied transverse stress of $I_c/I_{c0}$ at B = 16 T and B = 19 T for the 1-mm PIT wire. Solid and open symbols correspond to the measurement under load and after unload, respectively. The irreversible stress limit, denoted by dashed lines in the plot, depends on the intensity of the magnetic field.

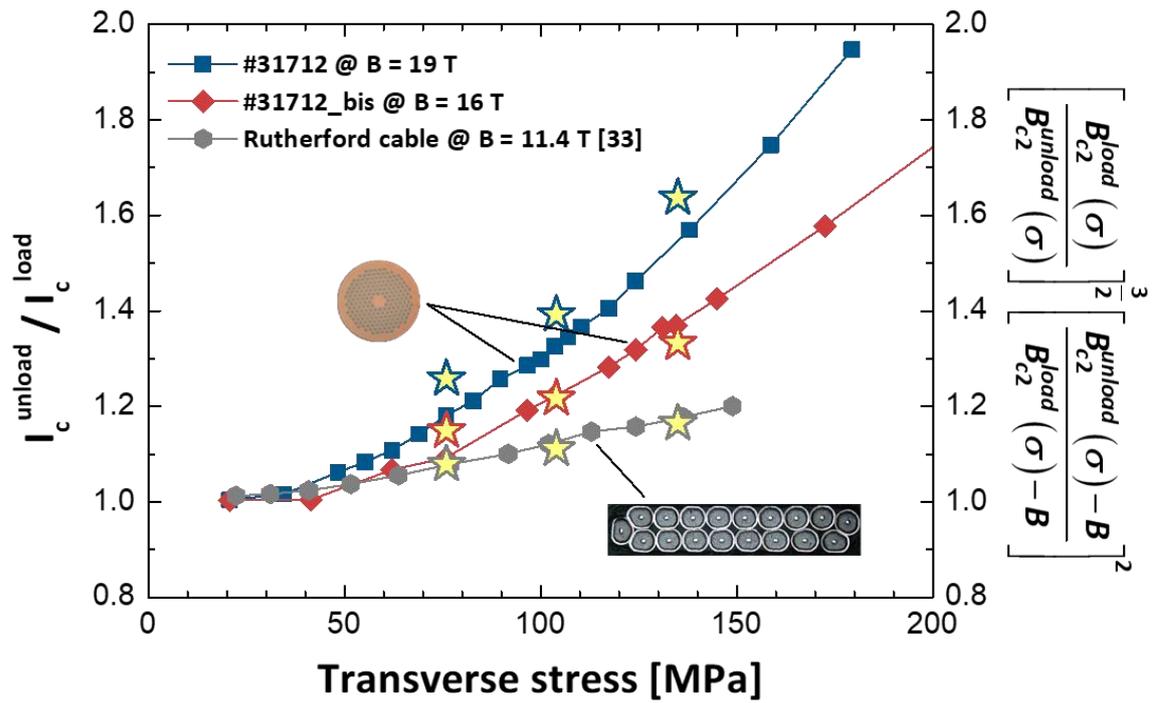

**Figure 4** Dependence on the applied transverse stress of $I_c^{unload}/I_{c0}^{load}$. Squares and diamonds correspond to the data of the single-wire tests in Fig. 3 at 19 T and 16 T, respectively. Hexagons result from a measurement performed on a Rutherford cable at 11.4 T and reported in Ref. [33]. The predictions of the model in eq. (3) are reported as open stars.